# Droplet Bouncing and Breakup during Impact on Microgrooved Surface

Laxman K. Malla[a], Nagesh D. Patil[b], Rajneesh Bhardwaj[b, *], Adrian Neild[c]

[a] IITB-Monash Research Academy, Indian Institute of Technology Bombay, Mumbai, 400076, India.

[b] Department of Mechanical Engineering, Indian Institute of Technology Bombay, Mumbai, 400076 India.

[c] Department of Mechanical and Aerospace Engineering, Monash University, Melbourne, VIC 3800, Australia.

[*]Corresponding author (Email: rajneesh.bhardwaj@iitb.ac.in, Phone: +91 22 2576 7534, Fax: +91 22 2576 6875)

*Abstract*

We experimentally investigate impact dynamics of a microliter water droplet on a hydrophobic microgrooved surface. The surface is fabricated using photolithography and high-speed visualization is employed to record the time-varying droplet shapes in transverse as well as longitudinal direction. The effect of the pitch of the grooved surface and Weber number on droplet dynamics and impact outcome are studied. At low pitch and Weber number, the maximum droplet spreading is found to be greater in the longitudinal direction than the transverse direction to the grooves. The preferential spreading inversely scales with the pitch at a given Weber number. In this case, the outcome is no bouncing (NB); however, this changes at larger pitch or Weber number. Under these conditions, the following outcomes are obtained as function of the pitch and Weber number - droplet completely bounces off the surface (CB), bouncing occurs with droplet breakup (BDB) or no bouncing due to Cassie to Wenzel wetting transition (NBW). In BDB and NBW, the liquid penetrates the grooves partially or completely beneath the droplet due to the wetting transition. The former results in droplet breakup alongside bouncing while the latter suppresses the bouncing. These outcomes are demarcated on Weber number – dimensionless pitch plane and the proposed regime map suggests the existence of a critical Weber number or pitch for the transition from one regime to the other. CB and BDB are quantified by plotting coefficient of restitution of the bouncing droplet and volume of daughter droplet left on the surface, respectively. The critical Weber number needed for the transition from CB to BDB is estimated using an existing mathematical model and is compared with the measurements. The comparison is good and provides insights in mechanism of the liquid penetration into the grooves. The present results on microgrooved surfaces are compared with published results on micropillared surfaces in order to assess water repelling properties of the two surfaces.

# 1 Introduction

In nature, several animals and plants utilize isotropic and anisotropic textured surfaces for drag reduction and self-cleaning. Examples include, anisotropic textures on shark skin or butterfly wing for low-drag locomotion. Whilst, a lotus leaf exhibits microscale isotropic structures on its surface for lower wettability. Outside the natural world, anisotropic wettability can be utilized to provide directional fluid transport in microfluidics and such textured surfaces can be used in engineering self-cleaning, low drag, de-icing and anti-fouling surfaces. Recent studies have demonstrated that the bioinspired microtextured surfaces exhibit drag reduction [1], self-cleaning [1] and super-repellency to water [2].

An impacting droplet on a solid surface spreads under the influence of kinetic energy and enabling a large deformation of the liquid-air interface. Gravitational energy helps in spreading if the droplet diameter is greater than the capillary length. The contact line motion is governed by dynamic wetting and also affects the spreading process [3, 4]. The spreading occurs if dynamic contact angle ($\theta$) is larger than advancing contact angle, $\theta \geq \theta_{adv}$. After reaching the maximum spreading, the droplet recoils due to dominance of surface energy over kinetic energy and the contact line recedes in this process. The receding occurs if the dynamic contact angle ($\theta$) is lesser than receding contact angle, $\theta \leq \theta_{rec}$. The contact line is pinned if $\theta_{rec} < \theta < \theta_{adv}$. The droplet bounces off the surface if the total energy of the droplet at the instance of the maximum receding exceeds combined initial surface energy and gravitational energy [5, 6]. The impact process is highly transient and the spreading time scales as $D_0/U_0$ [7], where $D_0$ is the droplet initial diameter and $U_0$ is the impact velocity. For instance, the time taken for a millimeter size droplet to spread on a hydrophillic surface with velocity lesser than 1 m s$^{-1}$ is on the order of milliseconds [3, 8]. The droplet impact dynamics on a microtextured surface is more complex and in general, due to lower surface wettability, the droplet bounces more readily off the surface. The lower wettability is attributed to Cassie-Baxter state (referred to as Cassie state, hereafter), in which air is trapped in the microtextures beneath the droplet. If liquid penetrates the microtextures, leaving no air pockets, it is known as Wenzel state. The penetration of the liquid is attributed to the lesser capillary pressure than the combined hammer and dynamic pressure (discussed later in section 3).

Several previous reports focused on the droplet impact dynamics on isotropic, micropillared surfaces and showed that impact velocity [9, 10], equilibrium contact angle [11], pitch of the pillars [9, 12–14], aspect ratio of the pillars [10], droplet volume [15, 16] influence the Cassie to Wenzel wetting transition. The transition may change the impact outcome from bouncing to no bouncing off the surface. Patil et al. [14] studied effect of pitch of the pillars and impact velocity, and proposed several droplet outcomes, such as, no bouncing, complete bouncing, and bouncing with droplet breakup on impact velocity-pitch plane. Recently, Liu et al. [17] and Liu and Kim [18] investigated the droplet bouncing on surfaces with conical pillars and cylindrical pillars with vertical overhangs, respectively. In the former study, the bouncing droplet diameter is comparable to the diameter at maximum spreading ("pancake bouncing") at Weber number ($We = \rho U_0^2 D_0/\gamma$), $We > 14.2$, where $\rho$, $U_0$, $D_0$ and $\gamma$ are density, impact velocity, initial droplet diameter and surface tension, repectively. While in the latter study [18], the surface exhibits superrepellency even to liquids with low surface tension. Very recently, Frankiewicz and Attinger [2] reported wettability characteristics of chemically etched copper surfaces with three tiers of roughness, analogous to a lotus leaf.

The wetting characteristic of anisotropic, microgrooved surfaces with rectangular grooves and ridges (shown in Figure 1 (a)) has been reported extensively. These surfaces exhibit anisotropic wetting characteristic i.e. the droplet spreads more in longitudinal direction than transverse direction to the grooves, as shown schematically in Figure 1 (b). In a notable paper, Chen et al. [19] reported smaller equilibrium contact angle ($\theta_{eq}$) in the longitudinal direction than in the transverse direction. Similar findings were reported by Zhao et al. [20] and Zhang and Low [21] for wetting behavior of a microgrooved surface and surface with two-tier microgrooves, respectively. Similarly, Kusumaatmaja et al. [22] found that the contact angle hysteresis is larger in the transverse direction than in the longitudinal direction for sub-micrometer scale corrugated polyimide surface. Bliznyuk et al. [23] reported that the preferential droplet spreading is more pronounced in Wenzel state than Cassie state. Ma et al. [24] utilized anisotropic silicon carbide surface and found that the anisotropic wetting increases with the ridge width while keeping the groove width constant.

In the context of droplet impact on the microgrooved surfaces, Kannan and Sivakumar [25] studied water droplet impact on a stainless steel rectangular grooved surface at $We = [50, 168]$ and found that the spreading liquid lamella changes in the presence of grooves. Pearson et al. [26] showed that ratio of wetted diameter in longitudinal direction to initial droplet diameter scales with Weber number for rectangular microgrooved surface for $We = [1, 500]$. Similarly, the spreading in the longitudinal direction increases with increase in Weber number on V-shape microgrooved surface for $We = [1, 100]$, as reported in Ref. [27]. The critical velocity for the Cassie to Wenzel wetting transition was found to be function of geometry and wetting characteristics of the base surface for a V-shape microgrooved surface [28]. Very recently, Yamamoto et al. [29] studied impact of a water droplet on surfaces on which parallel stainless steel razor blades were mounted with pitch of 100 µm and 250 µm. They reported droplet bouncing off the surface at $We = [0.04, 5]$ and the wetting transition at $We = (5, 10)$.

It is well-established by previous studies that the droplet preferentially spreads in longitudinal direction as compared to transverse direction on a microgrooved surface. However, most of the previous studies considered large Weber number in which the droplet is most likely in the Wenzel state. At low Weber number, the droplet is in Cassie state that may affect the preferential spreading. In particular, the pitch is expected to influence the preferential spreading since previous studies have shown dependence of surface wettability on the pitch in the Cassie state. The Cassie to Wenzel wetting transition may involve partial penetration of the liquid into the grooves that could result in breakup of the bouncing droplet. To the best knowledge, these effects have not been investigated thus far for the droplet impact dynamics on microgrooved surfaces. Therefore, the objective of the present paper is to investigate the combined effect of Weber number ($We$) and pitch ($p$) on the impact dynamics, anisotropic wetting characteristic, Cassie to Wenzel wetting transition, final impact outcome and droplet breakup during bouncing. We consider a microliter water droplet with Weber number, $We = [1.2, 20.4]$ on microgrooved surfaces with wide range of pitch, $p = [30, 76]$ µm.

The layout of the present paper is as follows. Section 2 and 3 describes experimental details and existing theory of Cassie to Wenzel wetting transition, respectively. We discuss the effect of pitch and Weber number on the impact dynamics in section 4.1 and 4.2, respectively. A comparison between theoretical Weber number of the wetting transition and present measurements is presented in section 4.2.1. A regime map summarizing different outcomes obtained in the measurements is discussed in section 4.3. We briefly discuss comparisons between measurements of the microgrooved and micropillared surface in section 4.4.

## 2   Experimental details

Microgrooved surfaces with different pitches ($p$, defined in Figure 1 (c), right) are fabricated using SU-8 photoresist on a silicon wafer and are coated with 10 nm platinum layer, using ultraviolet lithography. The details of the fabrication process can be found in our previous work [14]. SEM images of the top view of the surfaces with four pitches - $p = 30$ µm, 47 µm, 62 µm, and 76 µm - are shown in Figure 2 (a). Similarly, 3D views obtained from optical profilometer (Zeta-20, Zeta Instruments Inc) are shown in Figure 2 (b). The respective cross-sectional view (shown as 1-1' in top row of Figure 2 (b)) is shown in Figure 2 (c). The rectangular ridge width ($w$) and groove depth ($h$) are $22 \pm 2$ µm and $27 \pm 2$ µm, respectively, for all the surfaces.

The experimental setup shown in Figure 3 (a), consists of two high-speed cameras connected to a computer, a syringe with an adjustable stand to vary the height (or impact velocity) and two LED lamps. Microliter deionized water droplets are generated using the syringe with 31-gauge needle. The droplet diameter and range of impact velocity are 1.7±0.05 mm and [0.22, 0.92] m/s, respectively. The high-speed cameras are used to visualize the droplets in planes transverse and longitudinal to the direction of the grooves (Motion- Pro Y-3 classic camera in transverse view and Pixelink PL-C722MU-BL in longitudinal view). These directions are shown in Figure 1 (b). A long distance working objective (Qioptiq Inc.) is used with the two cameras. Two white LED lamps serve as a back light source. The magnification corresponds to 14 µm per pixel and 29 µm per pixel for the transverse and longitudinal view, respectively. The videos are recorded at 1500 and 750 frames per second (fps) in the transverse and longitudinal view, respectively.

Captive droplet method and sessile droplet evaporation are used to measure advancing contact angle ($\theta_{adv}$) and receding contact angle ($\theta_{rec}$), respectively. Equilibrium contact angle ($\theta_{eq}$) is measured for a gently deposited droplet on the surface, using the following equation considering a sessile spherical cap on the surface, $\theta_{eq} = 2\tan^{-1}(2H / D_{wetted})$ where $H$ and $D_{wetted}$ are sessile cap height and wetted diameter, respectively. The measured $\theta_{adv}$, $\theta_{rec}$, and $\theta_{eq}$ in the transverse and longitudinal directions along with contact angle hysteresis (CAH = $\theta_{adv}$ - $\theta_{rec}$) are plotted in Figure 3 (b) and 3 (c), respectively. The uncertainty in the contact angle measurements is ±5°. All experiments are performed at 25±2 ℃ and 50±5% relative humidity and are repeated three times to ensure repeatability.

## 3   Theory of Cassie to Wenzel wetting transition

We discuss theory reported in the literature for Cassie to Wenzel wetting transition. During the droplet impact, the interaction of the droplet with the microgrooved surface generates three pressures, namely, hammer pressure ($\Delta P_H$), dynamic pressure ($\Delta P_D$), and capillary pressure ($\Delta P_C$). $\Delta P_H$ is a thrust by the compressed liquid behind shock front generated due to the impact at a larger velocity on a rigid substrate and is expressed as follows [12, 30],

$$\Delta P_H = \alpha\rho CU_0 \tag{1}$$

where $\alpha$, $\rho$, $C$, and $U_0$ are an empirical coefficient, density of the droplet liquid, speed of the sound in the droplet liquid and impact velocity, respectively ($C$ = 1482 m/s in water). The coefficient, $\alpha$, is a fraction of the impact velocity, $U_0$, at which the compression wave travels through the droplet [30]. The dynamic pressure is given as follows:

$$\Delta P_D = 0.5 \rho U_0^{\,2} \qquad\qquad (2)$$

The capillary pressure is defined as the Laplace pressure acting along the air-liquid interface beneath the droplet and is expressed as follows [31]:

$$\Delta P_C = \gamma H \qquad\qquad (3)$$

where $\gamma$ and $H$ are surface tension and curvature of the interface, respectively. Since $\Delta P_H$ and $\Delta P_D$ act as wetting pressure, and $\Delta P_C$ acts as anti-wetting pressure [9, 10, 12, 32], a criterion for the liquid penetration inside the microgrooves i.e. Cassie to Wenzel wetting transition is given as follows:

$$\Delta P_H + \Delta P_D - \Delta P_C \geq 0 \qquad\qquad (4)$$

The wetting transition may occur by two mechanisms, namely, depinning or sagging [10, 33]. In the former, the contact line depins and the droplet slides downward on the sides of the ridge, as shown in Figure 1 (c, left) and capillary pressure is given in this case as follows [10],

$$\Delta P_C = \frac{-2\gamma \cos\theta_{adv}}{p - w} \qquad\qquad (5)$$

where $\theta_{adv}$, $p$, and $w$ are advancing contact angle on the base surface, pitch of the microgrooves and width of the ridge, respectively. In the sagging, curvature of the liquid-air interface increases due to larger wetting pressures and the interface touches the base surface, which leads to the wetting transition (Figure 1 (c, right)). The critical height of the microgroove required for the sagging is expressed as follows [34],

$$h_C = \frac{p - w}{2} \tan\left( \frac{\theta_{adv} - 90^o}{2} \right) \qquad\qquad (6)$$

# 4   Results and discussions

Results are presented for the impact of 1.7 mm water droplet on microgrooved surfaces with impact velocity and pitch in range of $U_0 = [0.22, 0.92]$ m/s and $p = [30, 76]$ μm, respectively, in the following subsections. The corresponding range of Reynolds number ($Re = \rho U_0 D_0/\mu$), Weber number ($We = \rho U_0^2 D_0/\gamma$) and dimensionless pitch ($p/D_0$) are $Re = [429, 1794]$, $We = [1.2, 20.4]$ and $p/D_0 = [0, 0.045]$, respectively, where $\rho$, $U_0$, $D_0$, $\mu$ and $\gamma$ are density, impact velocity, initial droplet diameter, dynamic viscosity and surface tension, respectively.

## 4.1 Effect of pitch

The wetting characteristics of the surfaces with different pitches are quantified in Figure 3 (b, c). The advancing, receding and equilibrium contact angles along with contact angle hysteresis (CAH) are plotted in transverse and longitudinal direction to the grooves in Figure 3 (b) and 3 (c), respectively. These directions are shown in Figure 1 (b). The increase in the angles with increase in the pitch implies decrease in surface wettability in Figure 3 (b, c). The angles and CAH are larger in the transverse direction than in the longitudinal direction, consistent with data of Kusumaatmaja et al. [22]. As explained by Chen et al. [19], the larger spreading in the longitudinal direction (or larger angles in the transverse direction) is due to the pinning of the contact line at the edge of the ridge while spreading in the transverse direction. This results in squeezing of the droplet in the transverse direction and stretching of the droplet in the longitudinal direction. These results show the anisotropy wetting characteristic of the microgrooved surfaces.

Figure 4 compares the droplet impact dynamics on flat and microgrooved surfaces with four different pitches, $p$ = 30 μm, 47 μm, 62 μm, 76 μm. The corresponding dimensionless pitch are $p/D_0$ = 0.018, 0.028, 0.037, and 0.045, respectively. The Weber number in all cases is $We$ = 6.5 that corresponds to impact velocity of $U_0$ = 0.52 m/s. The impact sequences obtained by high-speed visualization in the transverse direction are compared columnwise in Figure 4 and top insets shows the topology of the surface, recorded by SEM. The droplet initially spreads due to inertial and wetting forces ($t$ < 3.33 ms) and large deformation of the liquid-gas interface is observed. The surface forces cause the recoiling and the contact line recedes radially inwards. At $t$ = 11.33 ms, the droplet does not bounce on the flat as well as $p$ = 30 μm surface and bounces on the $p$ = 47 μm surface. The droplet bounces on the surfaces with pitch $p$ = 62 μm and $p$ = 76 μm, however, a small volume of the droplet is left on the surface due to droplet breakup (Figure 4). These three outcomes are referred to as – *no bouncing* (NB), *complete bouncing* (CB) and *bouncing with droplet breakup* (BDB), respectively, hereafter. The CB is explained due to lower wettability (or larger equilibrium contact angle, $\theta_{eq}$) at larger pitch (Figure 3 (b) and 3 (c)). In terms of the energies involved, the kinetic energy of the droplet converts into the surface energy until the maximum spreading and during recoiling the surface energy converts into the kinetic energy. If the sum of kinetic and surface energy exceeds the initial surface energy during recoiling, the droplet bounces off the surface [5, 6]. The measured bouncing times are in good agreement with the analytical model of Richard et al. [35] ($t \sim \sqrt{(\rho D_0^3/\gamma)}$ = 8.2 ms) and the measured wetted diameter ($D_{\text{wetted}}$) on the flat surface (~ 2.7 mm) and microgrooved surfaces (~ 2.5 mm) are consistent with prediction of an analytical model reported by Clanet et al. [36] ($D_{max} \sim D_0 We^{1/4} \sim 2.7$ mm).

The impact dynamics is further quantified by plotting the time-varying dimensionless wetted diameter $D_{wetted}^*$ in the transverse and longitudinal direction in Figure 5 (a) and 5 (b), respectively. The dimensionless wetted diameter and time are non-dimensionalized as $D_{wetted}^* = D_{wetted} / D_0$ and $t^* = tU_0/D_0$, respectively and superscript * denotes the dimensionless variable. $D_{wetted}^*$ is compared for surfaces with different pitches and it increases to a maxima during initial spreading and further decreases due to the receding in all cases at $t^* < 3$. The maximum $D_{wetted}^*$ corresponds to the flat surface due to its largest wettability (Figure 3 (b, c)). $D_{wetted}^*$ attains a plateau value after the droplet becomes sessile on the surface for NB at $t^* > 3$. In Figure 5, $D_{wetted}^* = 0$ and $D_{wetted}^* < 0.3$ denote CB and BDB, respectively. In cases of BDB, $D_{wetted}^*$ stays almost constant for $3 < t^* < 12$ and is slightly larger for $p = 76$ μm surface as compared to $p = 62$ μm surface. The latter is due to the fact that a larger pitch surface captures a larger volume of water into the grooves during Cassie to Wenzel wetting transition (discussed in detail in section 4.2.1). $D_{wetted}^*$ starts increasing after the droplet reimpacts for CB and BDB at around $t^* = 6$ and 11, respectively (Figure 5).

The maximum spreading is larger in the longitudinal direction than in the transverse direction in Figure 5 (a) and 5 (b), which is attributed to anisotropic wetting charateristic of the surface. In order to quantify this, we define percentage change in the maximum wetted diameter in the longitudinal direction with respect to that in the transverse direction ($\eta$) at a given dimensionless pitch ($p/D_0$) as follows,

$$\eta(\%) = \frac{D_{wetted,max,longitudinal} - D_{wetted,max,transverse}}{D_{wetted,max,transverse}} \times 100 \qquad (7)$$

Figure 6 plots the variation of $\eta$ with dimensionless pitch ($p/D_0$) at $We = 6.5$. We note that the spreading is considerably larger in the longitudinal direction in all cases and $\eta = 27\%$, 22%, 19%, and 17% for $p/D_0 = 0.018$, 0.028, 0.037, and 0.045, respectively.

## 4.2    Effect of Weber number

The effect of the Weber number ($We$) is investigated by varying it in range of $We = [1.2, 20.4]$ on a $p = 47$ μm surface ($p/D_0 = 0.028$). The corresponding range of the impact velocity is $U_0 = [0.22, 0.92]$ m/s. The images acquired by the high-speed visualization in Figure 7 show NB for $We = 2.8$ ($U_0 = 0.34$ m/s) and $We = 4.5$ ($U_0 = 0.43$ m/s), CB for $We = 6.5$ ($U_0 = 0.52$m/s) and *We* = 9.3 ($U_0$ = 0.62 m/s), and BDB for *We* = 16.2 ($U_0$ = 0.82 m/s). In all cases, most of the initial kinetic energy converts into surface energy upon impact whilst some is also dissipated during the spreading processes due to viscosity [35]. As such, a decrease in the initial kinetic energy gives rise to an increase in the surface energy stored, which in turn aids the receding of the droplet after spreading. As explained earlier, CB and BDB occurs if the sum of kinetic and surface energy exceeds the initial surface energy during recoiling.

The corresponding time-variation of $D^*_{\text{wetted}}$ in transverse and longitudinal direction for the *p* = 47 μm surface is shown in Figure 8 (a) and 8 (b), respectively. As expected, the instance of maximum spreading inversely scales with impact velocity ($t_{spreading} \sim D_0 / U_0$ [7] or $t^*_{spreading} = t_{spreading} U_0 / D_0 \sim 1$). A larger impact velocity results in larger spreading as well as receding of the contact line, as shown in Figure 8. In cases of CB ($D^*_{\text{wetted}} = 0$), the droplet remains in air for longer time for larger Weber number (or impact velocity). In Figure 8 (b), we note oscillations of $D^*_{\text{wetted}}$ in the longitudinal direction after completion of the recoiling for NB and after reimpact of the droplet for CB and BDB. These oscillations are attributed to the conversion of kinetic and surface energy into each other [6] and were also observed for the impact of a water droplet on glass [3]. The amplitude of the oscillation decays with time due to viscous dissipation of the energies and eventually the droplet reaches to a sessile state. The oscillations are absent in the transverse direction since contact line pins at the microgroove edge i.e. anisotropic wetting of the surface.

We further quantify the anisotropic wetting as function of the Weber number (*We*) and dimensionless pitch ($p/D_0$). Figure 9 plots the percentage change in the maximum wetted diameter in the longitudinal direction with respect to that in the transverse direction ($\eta$, eq. 7) as function of the dimensionless pitch ($p/D_0$) for different cases of *We*. At low *We*, $\eta$ is larger and decreases with $p/D_0$. This is due to the pinning of the contact line at the edge of the ridge, which causes obstruction for the droplet to spread in the transverse direction. The chances of the obstruction are larger at lower $p/D_0$ because of presence of higher number of the ridges beneath the droplet. However, $\eta$ stays constant (~ 20 - 30%) at larger *We* because the kinetic energy dominates over the surface energy at larger *We* and consequently, the preferential spreading in the longitudinal direction suppresses, as seen in Figure 9. The largest $\eta$, $\eta$ = 80%, is obtained for $p/D_0$ = 0.018 (*p* = 30 μm), *We* = 1.2 ($U_0$ = 0.22 m/s).

### 4.2.1 Cassie to Wenzel wetting transition

We examine the Cassie to Wenzel wetting transition on the microgrooved surfaces in this section. As shown in Figure 10, the droplet impact on $p = 76$ μm surface results in NB at $We = 1.2$ ($U_0 = 0.22$ m/s), CB at $We = 2.8$ ($U_0 = 0.34$ m/s), $We = 4.5$ ($U_0 = 0.43$ m/s), BDB at $We = 6.5$ ($U_0 = 0.52$ m/s), $We = 9.3$ ($U_0 = 0.62$ m/s) and again no bouncing at $We = 16.2$ ($U_0 = 0.82$ m/s). Therefore, a critical Weber number (or impact velocity) exists at which the droplet outcome changes from CB to BDB and the liquid penetrates into the grooves. The no bouncing at $We = 16.2$ ($U_0 = 0.82$ m/s) is due to the Cassie to Wenzel wetting transition and is referred to as – *no bouncing due to the wetting transition* (NBW). The time-varying droplet shapes for NBW are plotted in Figure 11 (a). The microgrooves are visible in Figure 11 (a) before the impact at $t = 0$ ms and air-filled and liquid-filled microgrooves are visible beneath the droplet at 1.33 ms. In addition, capillary wave propagation is visible on the liquid-gas interface at 1.33 ms. From 4.67 ms to 6.67 ms, the contact line recedes and the droplet recoils, however, it sticks in the microgrooves due to the penetration of water into the grooves and does not bounce. Whereas, for the CB case on $p = 76$ μm surface at $We = 4.5$ ($U_0 = 0.43$ m/s) in Figure 11 (b), the microgrooves beneath the droplet are completely filled with air during spreading at $t = 2.67$ ms, receding at $t = 6.00$ ms and bouncing at $t = 9.33$ ms in Figure 11 (b). The droplet remains in Cassie state even after it impacts again at 44.67 ms.

We further calculate the critical $We$ required for the wetting transition using the existing theory presented in section 3 and compare it with our measurements for three cases of the pitch, $p = 47$ μm, 62 μm and 76 μm, corresponding to dimensionless pitch, $p/D_0 = 0.028$, 0.037, and 0.045. As discussed in section 3, the droplet penetrates inside the microgrooves if the combined wetting pressure ($\Delta P_H + \Delta P_D$) exceeds the capillary pressure ($\Delta P_C$). The critical height estimated using eq. 6 for the wetting transition due to the sagging of the interface (Figure 1 (c, right)) for $p = 47$ μm, 62 μm and 76 μm is on the order of 1 μm. Since the height of the ridge is 27 μm, the wetting transition occurs by depinning of the contact line (Figure 1 (c, left)) in the present measurements. The critical $We$ is estimated using dimensionless form of eq. 4, expressed as follows,

$$\Delta P_H^* + \Delta P_D^* = \Delta P_C^* \tag{8}$$

where superscript * denotes dimensionless variable normalized with $\rho U_0^2$. The coefficient $\alpha$ in the expression of the dimensionless hammer pressure ($\Delta P_H^* = \alpha C / U_0$, dimensionless form of eq. 1) is a function of impact velocity and reported values of $\alpha$ in the literature are listed in

Table 1. Extrapolating the dependence of $\alpha$ on $U_0$ using data in Table 1, the estimated range of $\alpha$ is [$10^{-5}$, $10^{-4}$] for $U_0$ = [0.22, 0.92] m/s in the present work. We plot $\Delta P_H^* + \Delta P_D^*$, and $\Delta P_C^*$ using the respective equations given in section 3 as function of $We$ for the surfaces with the following pitches, $p$ = 47 μm ($p/D_0$ = 0.028), $p$ = 62 μm ($p/D_0$ = 0.037) and $p$ = 76 μm ($p/D_0$ = 0.045) in Figure 12 (a), 12 (b), and 12 (c), respectively. The intersection of plots of $\Delta P_H^* + \Delta P_D^*$ and $\Delta P_C^*$ is shown by a filled circle in Figure 12 (a-c) and it represents the theoretical estimate of $We$ needed for the wetting transition. $\alpha$ = $7 \times 10^{-5}$ is used as a fitting parameter in present calculations for $\Delta P_H^*$ in Figure 12 (a-c) and this value lies in the estimated range of $\alpha$ mentioned earlier. The measured minimum $We$ for the CB and BDB are shown by green and blue vertical lines, respectively, in Figure 12 (a-c). Since the filled circle lies between the vertical lines for three cases of the pitch in Figure 12, the calculated values of the critical $We$ are consistent with the measurements. These results show that the critical $We$ required for the wetting transition inversely scales with the pitch.

### 4.2.2 Bouncing with droplet breakup

As discussed earlier in section 4.1, BDB occurs due to partial penetration of water in the microgrooves (Cassie to Wenzel wetting transition) at large pitch and large Weber number. Figure 13 (a) shows time-varying droplet shapes on $p$ = 62 μm surface ($p/D_0$ = 0.037) with $We$ = 16.2 ($U_0$ = 0.82 m/s). The droplet spreads (0 ms to 5.33 ms), recedes (10 ms to 11.33 ms) and capillary wave propagation on the liquid-gas interface is noted after 8.67 ms. Due to the penetration of water in the microgrooves, the droplet sticks to the base surface and capillary wave propagation on the liquid-gas interface causes necking of the interface (8 ms to 12.67 ms). This results in the breakup of the droplet at 13.33 ms. We further quantify volume of daughter droplet left on the surface ($V$) from the recorded images. An axisymmetric droplet is assumed in the image in order to calculate volume. The percentage of $V$ with respect to initial droplet volume is plotted in Figure 13 (b) on Weber number ($We$) - dimensionless pitch ($p/D_0$) plane for the BDB ($V$ = 0 for NB, CB and NBW). The contour plot shows that the maximum $V$ % corresponds to $p/D_0$ = 0.037 surface at $We$ = 16.2. Therefore, there exists an optimal pitch as well as Weber number for achieving the maximum $V$.

### 4.3 Regime map

A regime map is proposed using present measurements for different impact outcomes. Figure 14 (a) shows the outcomes, namely, NB, CB, BDB, and NBW, on Weber number ($We$) - dimensionless pitch ($p/D_0$) plane. Figure 14 (a) shows NB is observed in all cases of *We* at $p/D_0$ = 0.018 ($p$ = 30 μm). As *We* increases at $p/D_0$ = 0.028 ($p$ = 47 μm), the outcome changes from NB to CB and CB to BDB. Similarly, at $p/D_0$ = 0.037 ($p$ = 62 μm) and $p/D_0$ = 0.045 ($p$ = 76 μm), the outcome changes from NB to CB, CB to BDB and BDB to NBW. Therefore, there exists a critical *We* at a constant pitch or a critical pitch at a constant *We* for change from one regime to other. The critical *We* is the lowest for NB to CB at a given pitch. At larger *We*, CB to BDW and BDW to NBW occurs due to Cassie to Wenzel wetting transition, as explained in section 4.2.1.

In order to characterize the intensity of the bouncing, we estimate coefficient of restitution ($\varepsilon$) for CB or BDB cases. In this context, Richard and Quéré [37] plotted coefficient of restitution for the bouncing droplets on non-textured superhydrophobic surfaces. It is defined as $\varepsilon = U/U_0$, where $U$ and $U_0$ are velocity of the droplet just after bouncing and impact velocity, respectively. $U$ is estimated as $\sqrt{2gh}$, where $h$ is the maximum height of the center of gravity of the droplet achieved after the bouncing. In BDB cases, $h$ is measured after neglecting the daughter droplet that remains on the surface.

The contours of $\varepsilon$ on *We* - ($p/D_0$) plane are plotted in Figure 14 (b), that shows $\varepsilon$ is the maximum for the CB cases at the largest pitch ($p/D_0$ = 0.045) and at *We* = [2.8 - 4.5]. Further increase in *We* at this pitch results in the penetration of liquid into the microgrooves and the outcome is BDB that lowers $\varepsilon$, as plotted in Figure 14 (b). At *We* = [2.8 - 4.5], $\varepsilon$ slightly decreases on $p/D_0$ = 0.037 surface in comparison to $p/D_0$ = 0.045 surface due to slightly larger surface wettability of the former than the latter (Figure 3 (b, c)). At $p/D_0$ = 0.028, the droplet bouncing is first observed at *We* = 6.5 with lower $\varepsilon$ (~ 0.3) because it has larger wettability in comparison to surfaces with $p/D_0$ = 0.037 and $p/D_0$ = 0.045. In addition, the droplet experiences more number of ridges beneath it for $p/D_0$ = 0.028, which leads to more dissipation of the kinetic energy (which lowers $\varepsilon$) due to pinning and depinning of advancing or receding contact line [27]. Therefore, the coefficient of restitution is the largest for CB at the largest pitch considered and at an optimal Weber number.

## 4.4    Comparison with results on micropillared surfaces

We compare present results on the microgrooved surfaces with previous results on micropillared surfaces [14]. The height and width of the pillar in Ref. [14] is almost equal to that of the ridge in the present study (within ±2 μm). As explained in section 4.1, the spreading on the microgrooved surface is more in longitudinal direction than in transverse direction to the grooves. However, the spreading is isotropic on micropillared surfaces [14]. Further, we compare minimum Weber number ($We_{min}$) required for regimes; CB, BDB and NBW, as function of dimensionless pitch ($p/D_0$), in Figure 15 (a), 15 (b) and 15 (c), respectively. Comparisons show that $We_{min}$ for CB, BDB and NBW is larger for the microgrooved surface as compared to the micropillared surface at a given pitch. This is explained by the pinning and depinning of the contact line at edge of the ridge during spreading that dissipates larger kinetic energy in the former. Therefore, a larger Weber number is required to achieve CB, BDB and NBW on the microgrooved surface. Regarding comparison between water repelling properties of the two surfaces, the microgrooved surface is a better choice at larger pitch since Weber number (or impact velocity) required to achieve NBW is larger. Similarly, the micropillared surfaces perform well at lower pitch because the Weber number required to achieve CB is lower.

## 5    Conclusions

We investigate impact dynamics of a microliter water droplet (1.7 mm diameter) on a hydrophobic microgrooved surface, manufactured using photolithography. Time-varying droplet shapes are obtained in longitudinal and transverse direction to the grooves using high-speed visualization. The range of impact velocity and pitch of the microgrooves are $U_0 = [0.22, 0.92]$ m/s and $p = [30, 76]$ μm, respectively. The corresponding range of Reynolds number, Weber number and dimensionless pitch are $Re = [429, 1794]$, $We = [1.2, 20.4]$ and $p/D_0 = [0, 0.045]$, respectively. The surfaces exhibit anisotropic wetting characteristic and measured advancing, receding, and equilibrium contact angles are around 20 percent larger in the transverse direction. The droplet spreads more in longitudinal direction and this preferential spreading at low Weber number inversely scales with the pitch. The largest percentage increase in the maximum wetted diameter in the longitudinal direction with respect to that in the transverse direction is around 80% at $We = 1.2$ at $p/D_0 = 0.018$.

The following regimes are obtained with increasing pitch at constant Weber number or with increasing Weber number at constant pitch: no bouncing (NB) → complete bouncing (CB) → bouncing with droplet breakup (BDB) → no bouncing due to Cassie to Wenzel wetting transition (NBW). A regime map is proposed on Weber number ($We$)-dimensionless pitch ($p/D_0$) plane for demarcations of the impact outcomes. BDB and NBW occur due to partial and full penetration of water into the microgrooves during Cassie to Wenzel wetting transition, respectively. We quantify the volume of daughter droplet left on the surface after the droplet breakup for BDB and the maximum volume corresponds to an optimal pitch and $We$. The coefficient of restitution for CB and BDB cases is plotted on *We* - *p/D*$_0$ plane and the largest coefficient corresponds to the largest pitch and optimal *We* for CB.

We compare measured critical *We* needed for the wetting transition with an existing mathematical model, based on the balance of capillary pressure, dynamic pressure and water hammer pressure on the liquid-gas interface across the groove. The comparisons are good in all cases of pitch and results suggest that the critical *We* inversely scales with the pitch. The depinning of the contact line is found to be the possible mechanism for the wetting transition in the present measurements. Comparison between results on microgrooved and micropillared surfaces show that the minimum *We* required to achieve CB, BDB, and NBW at a given pitch is larger for the former. Hence, the former exhibit better water repelling properties than the latter at larger pitch while the latter is a better choice at lower pitch. The present results provide fundamental insights into droplet impact dynamics on microgrooved surfaces that may help to design technical applications such as self-cleaning and low-drag textured surfaces.

# 6   Associated Content

The following supplementary information is provided:

- High-speed visualization of impact dynamics in transverse direction for 1.7 mm water droplet on surfaces with several pitches at Weber number, *We* = 6.5 (S1.avi)
- High-speed visualization of impact dynamics in transverse direction for 1.7 mm water droplet on surface with pitch, *p* = 47 μm at several Weber numbers (S2.avi)

# 7   Acknowledgments

R.B. gratefully acknowledges financial support by an internal grant from Industrial Research and Consultancy Centre, IIT Bombay. The microgrooved surfaces were fabricated at Centre of Excellence in Nanoelectronics and SEM images were recorded at Centre for Research in Nanotechnology and Science, IIT Bombay.

# 9  Tables

Table 1. Reported values of the coefficient ($\alpha$) in eq. 1 for estimating water hammer pressure during impact of a water droplet on different substrates.

| Study by | Substrate | Impact velocity | $\alpha$ |
|---|---|---|---|
| Engel, 1955 [30] | Glass | 4.84 m/s | 0.2 |
| Deng et al., 2009 [12] | Silicon | 3 m/s | 0.2 |
| Kwon and Lee, 2012 [13] | SU-8 | 1.2 m/s – 3.1 m/s | 0.003 |
| Quan and Zhang, 2014 [38] | SU-8 | 1.89 m/s | 0.003 |

# 10 Figures

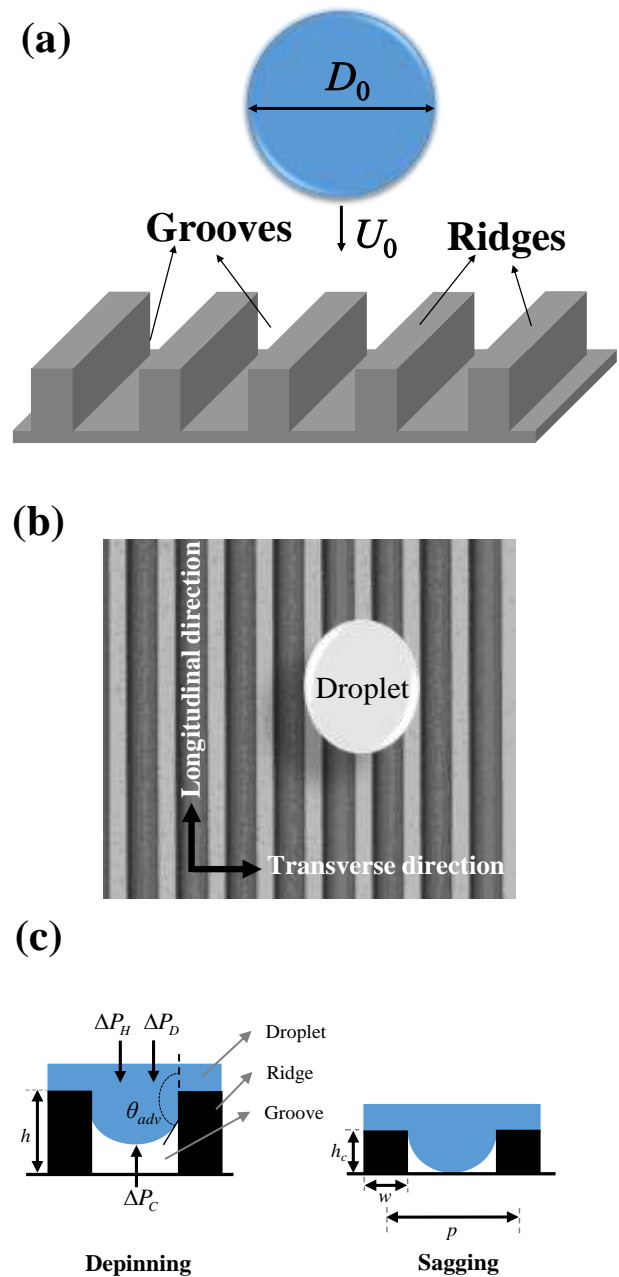

Figure 1: (a) Schematic of a droplet of initial diameter $D_0$ impacting with velocity $U_0$ on a microgrooved surface with rectangular ridges. (b) Top view of a sessile droplet on the surface. Light and dark colored patches are ridges and grooves, respectively. (c) Droplet penetration into the microgrooves could occur by depinning of the contact line (left) or by sagging of the liquid-air interface (right) [10, 33].

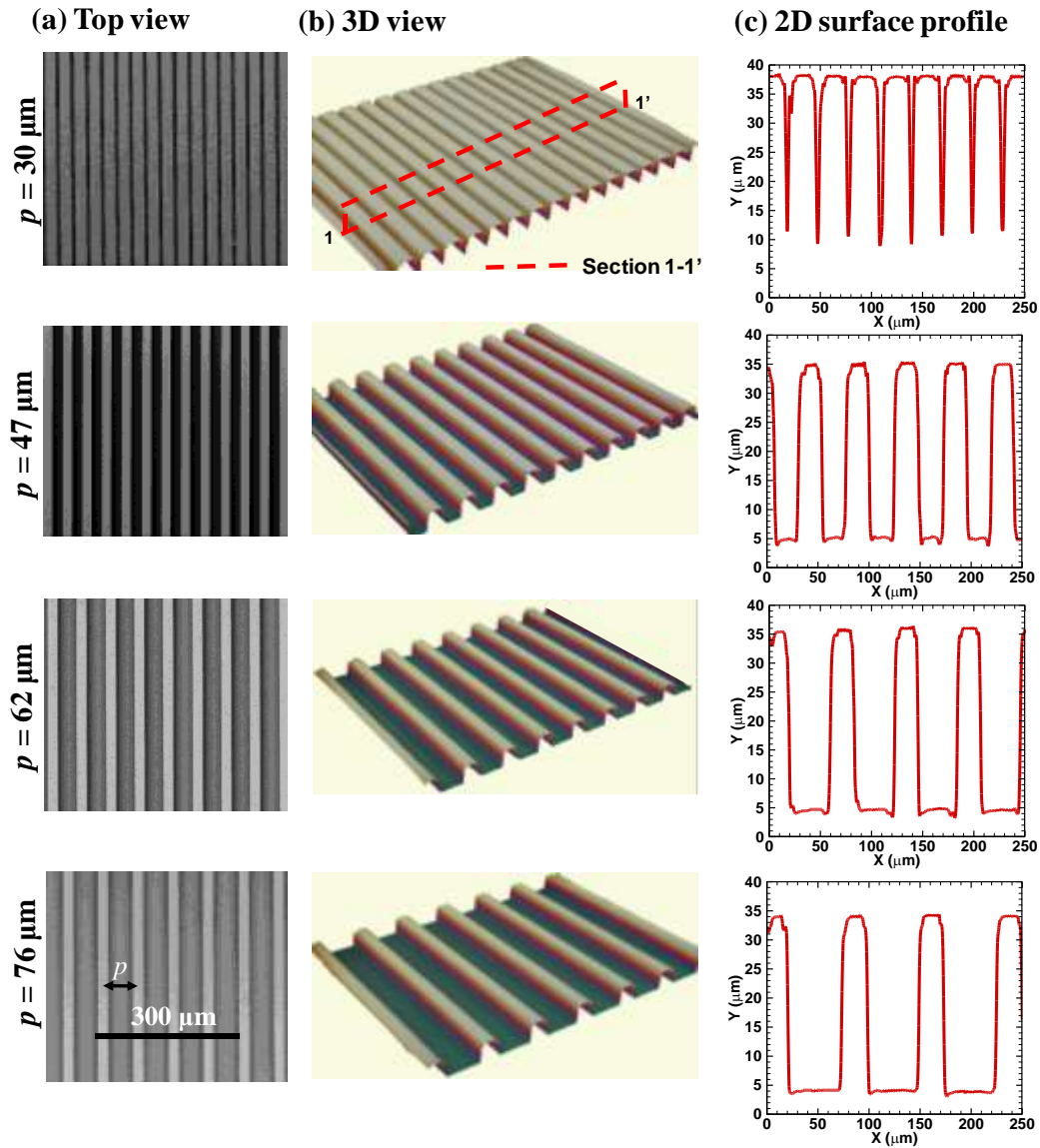

Figure 2: Characterization of microgrooved surfaces of different pitches ($p$). (a) Top view of the surfaces obtained by SEM. Scale is shown in image plotted in last row and first column. (b) 3D view of the surface obtained by optical profilometer. (c) 2D cross-sectional profiles across plane 1-1′ obtained using 3D profilometer data. The cross-section plane 1-1′ is shown in first row and second column.

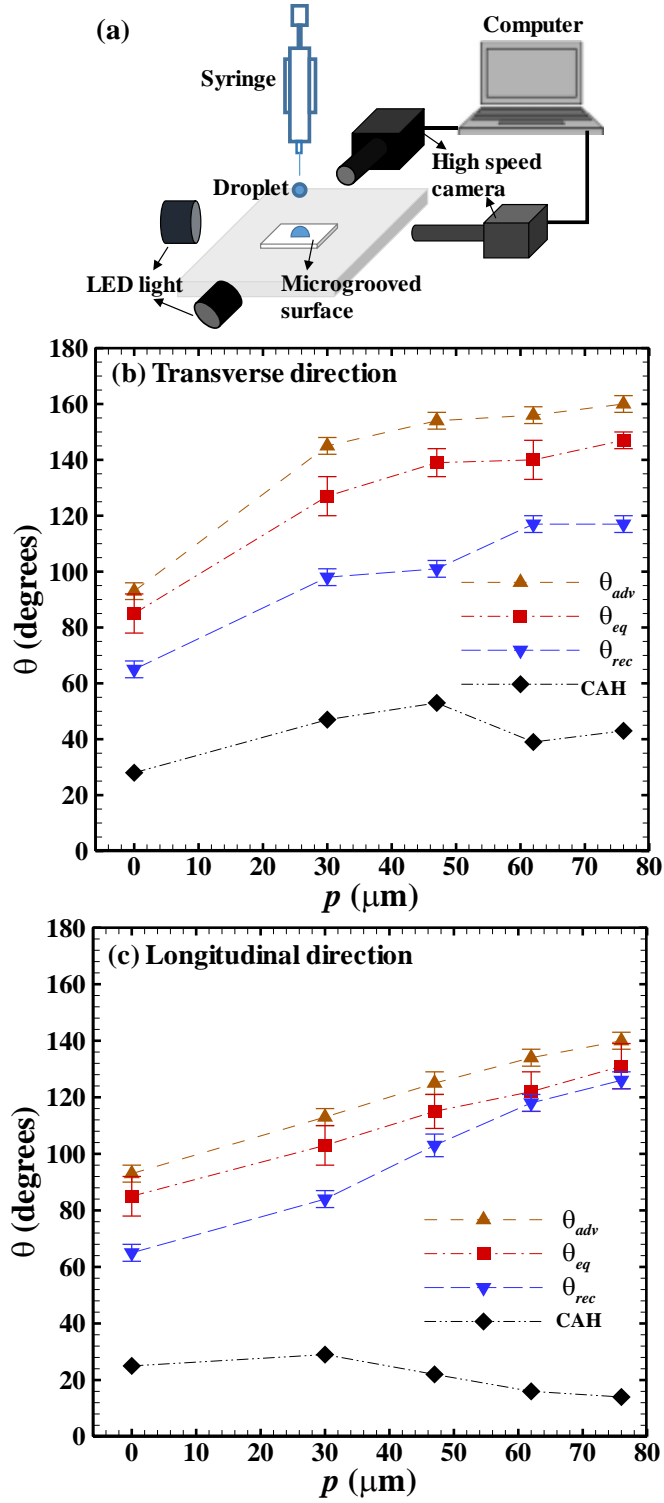

Figure 3: (a) Schematic of the experimental setup. (b) Measured advancing contact angle ($\theta_{adv}$), equilibrium contact angle ($\theta_{eq}$), receding contact angle ($\theta_{rec}$) and contact angle hysteresis (CAH = $\theta_{adv}$ - $\theta_{rec}$) for flat and microgrooved surfaces of various pitches ($p$) in the transverse direction. (c) Corresponding angles and hysteresis in the longitudinal direction.

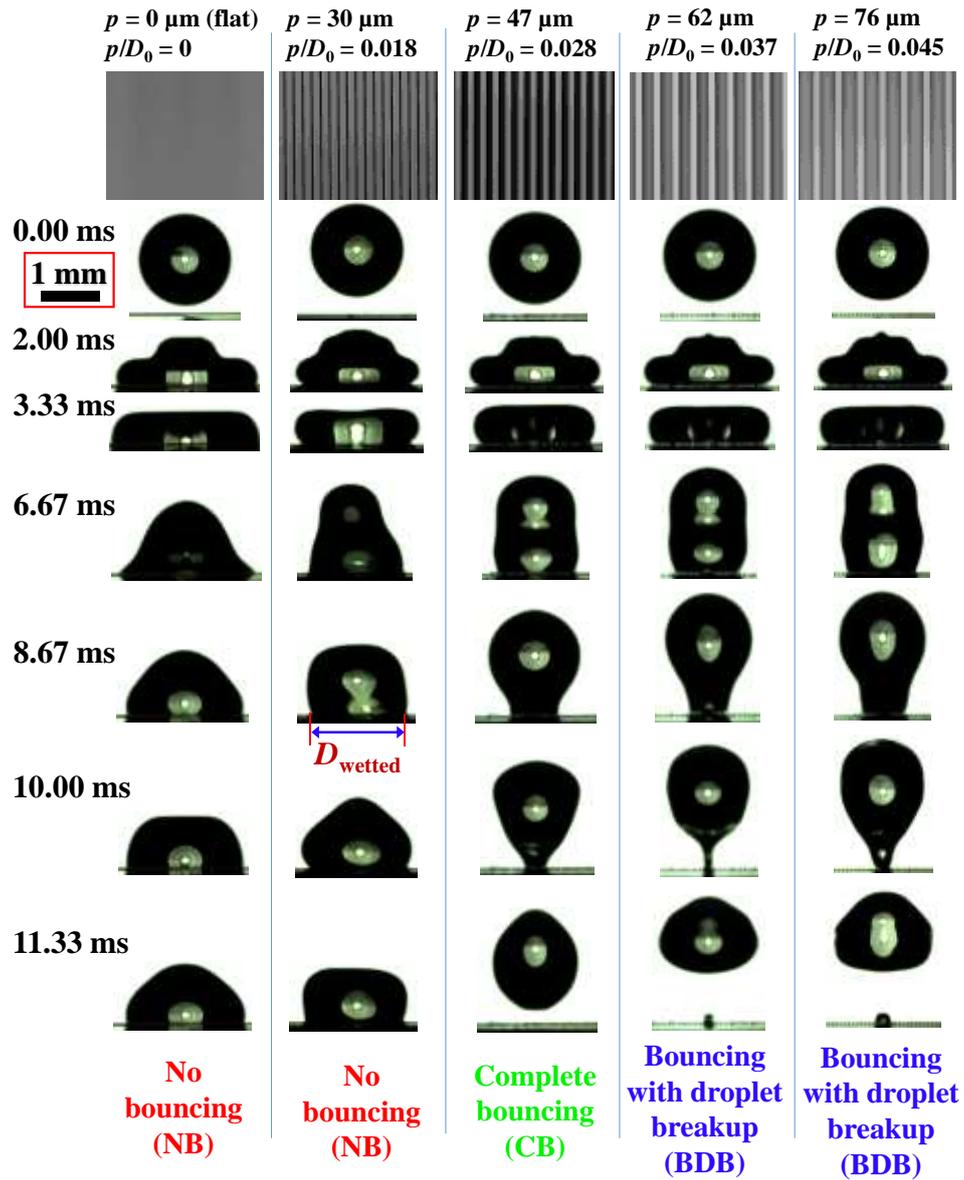

Figure 4: Image sequence obtained by high-speed visualization in transverse direction for impact of a microliter water droplet of 1.7 mm diameter on a flat and microgrooved surfaces of various pitches ($p$) at Weber number, $We = 6.5$ ($U_0 = 0.52$ m/s). Columns show different cases of pitch and scale is shown at top-left. See also the associated supplementary movie, S1.avi.

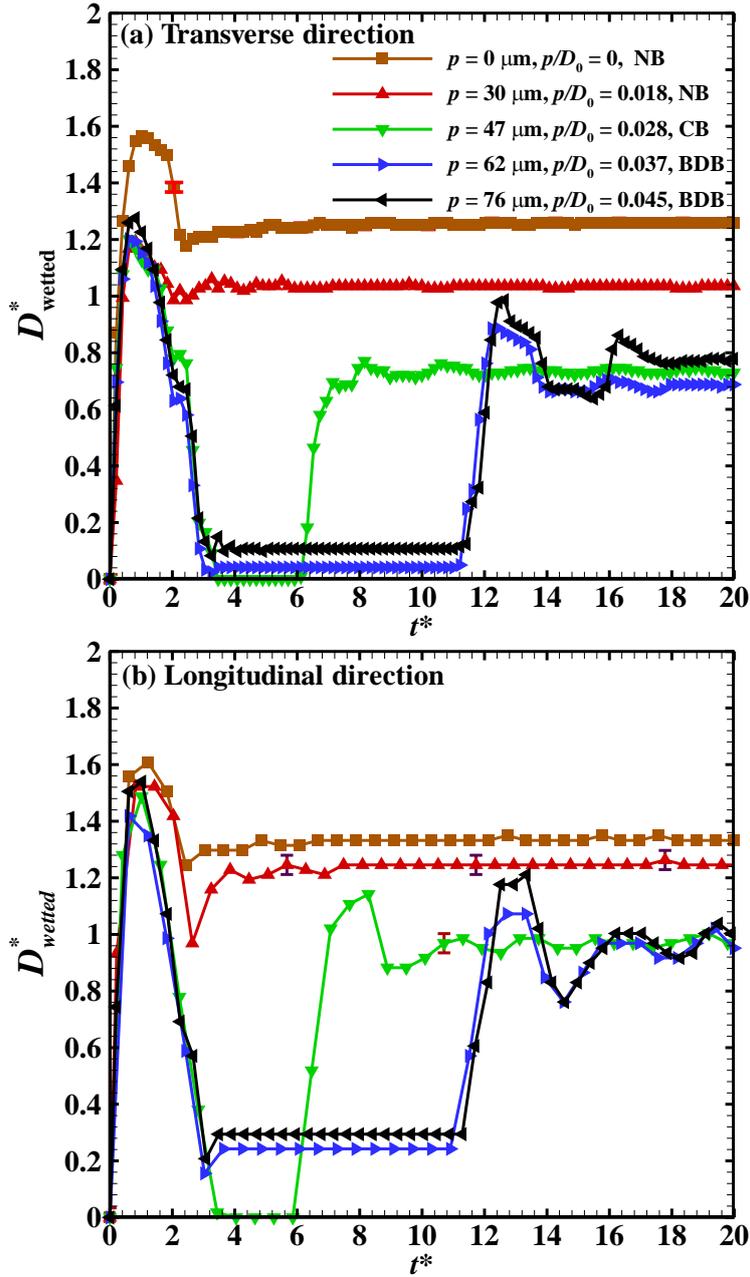

Figure 5: Time-varying dimensionless wetted diameter ($D^*_{\text{wetted}}$) for water droplet impact on flat surface and microgrooved surfaces of various pitches ($p$) at Weber number, $We = 6.5$ ($U_0 = 0.52$ m/s) in transverse (a) and longitudinal (b) direction. Different outcomes are obtained, namely, no bouncing (NB), complete bouncing (CB) and bouncing with droplet breakup (BDB). Temporal resolution in the longitudinal direction is half to that in the transverse direction. Only few error bars are shown for clarity.

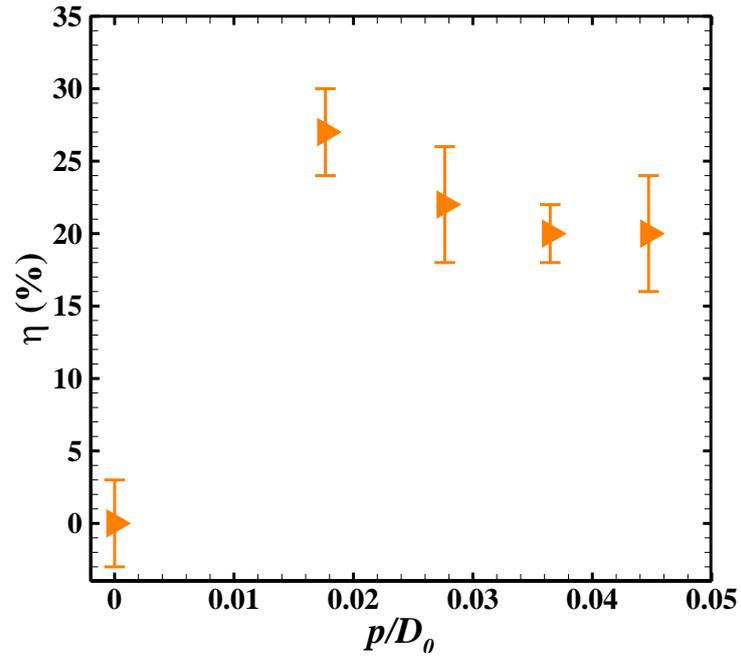

Figure 6: Percentage increase in the maximum spreading in longitudinal direction with respect to that in the transverse direction ($\eta$, eq. 7) as function of dimensionless pitch ($p/D_0$) for Weber number, $We = 6.5$ ($U_0 = 0.52$ m/s).

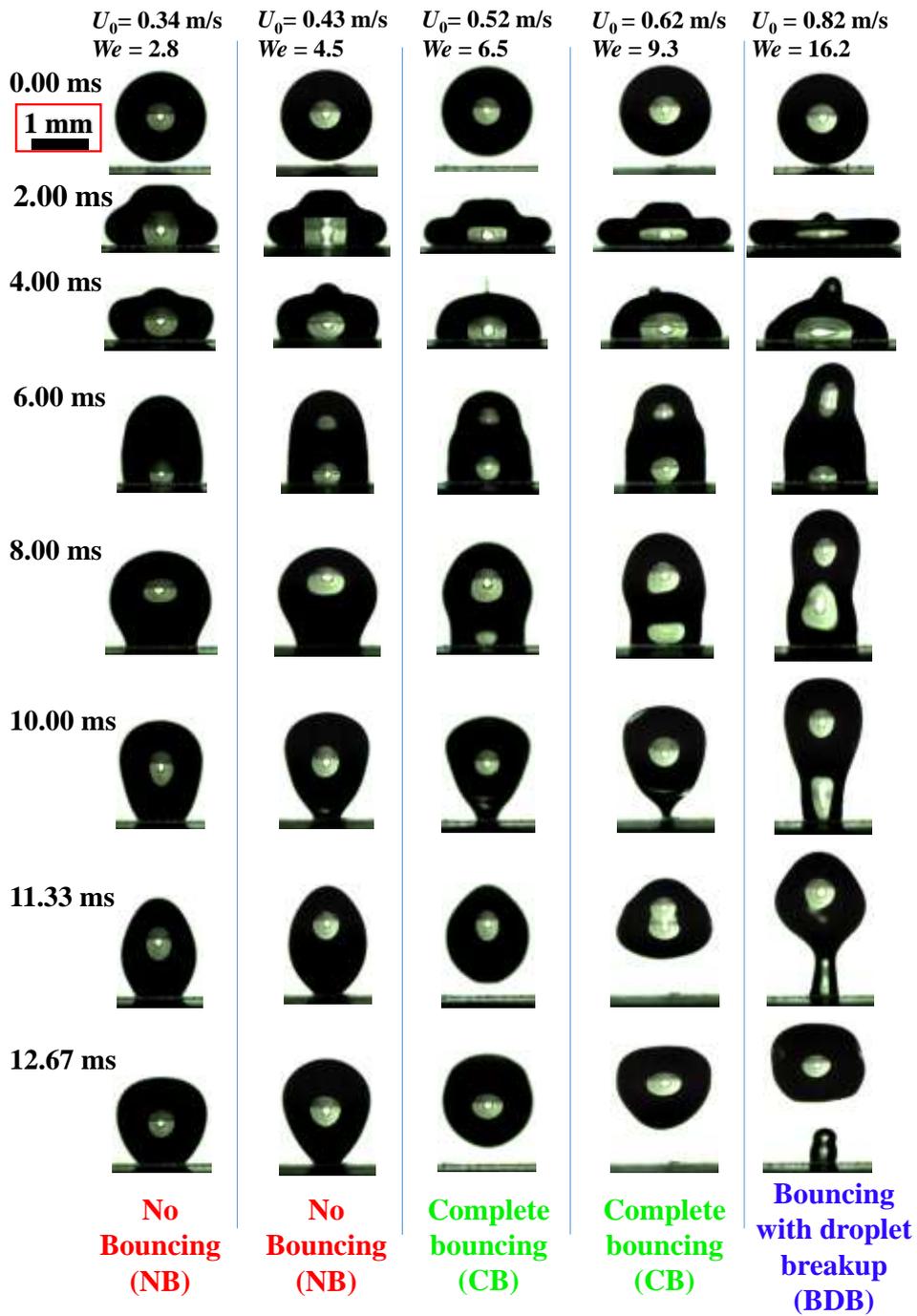

Figure 7: Image sequence obtained by high-speed visualization in transverse direction during impact of a microliter water droplet of 1.7 mm diameter on a microgrooved surface of pitch, *p* = 47 µm. Columns show different cases of Weber number (or impact velocity) and scale is shown on top-left.

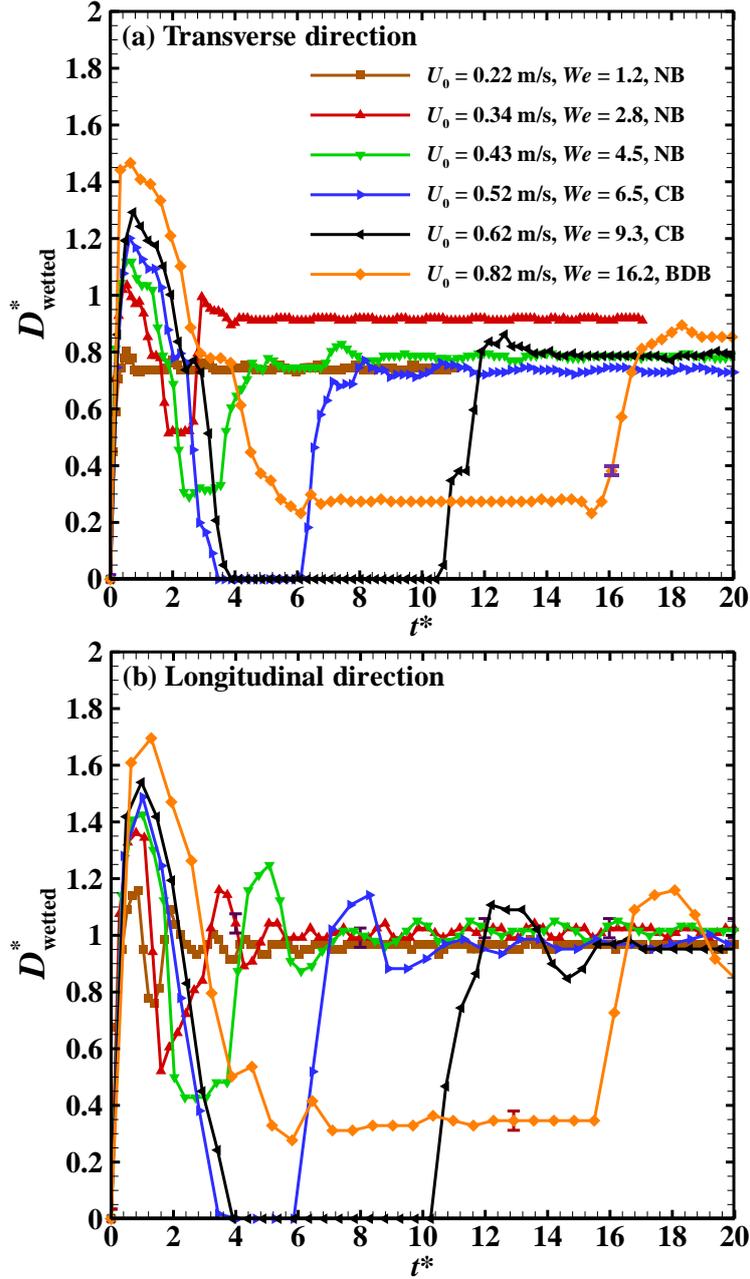

Figure 8: Time-varying dimensionless wetted diameter ($D^*_{\text{wetted}}$) of water droplet impact on a pitch surface of 47 μm with various Weber numbers (or impact velocities) in transverse (a) and longitudinal (b) direction. Different outcomes are obtained, namely, no bouncing (NB), complete bouncing (CB) and bouncing with droplet breakup (BDB). Temporal resolution in the longitudinal direction is half to that in transverse direction. Only few error bars are shown for clarity.

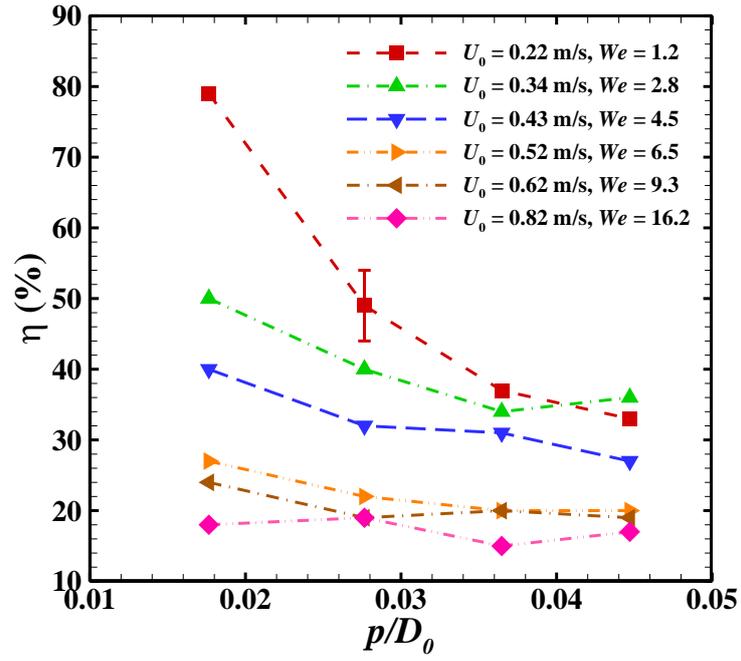

Figure 9: Percentage increase in the maximum spreading in longitudinal direction with respect to that in transverse direction ($\eta$, eq. 7) as function of dimensionless pitch ($p/D_0$) for different cases of Weber number (or impact velocity). Only one error bar is shown for clarity.

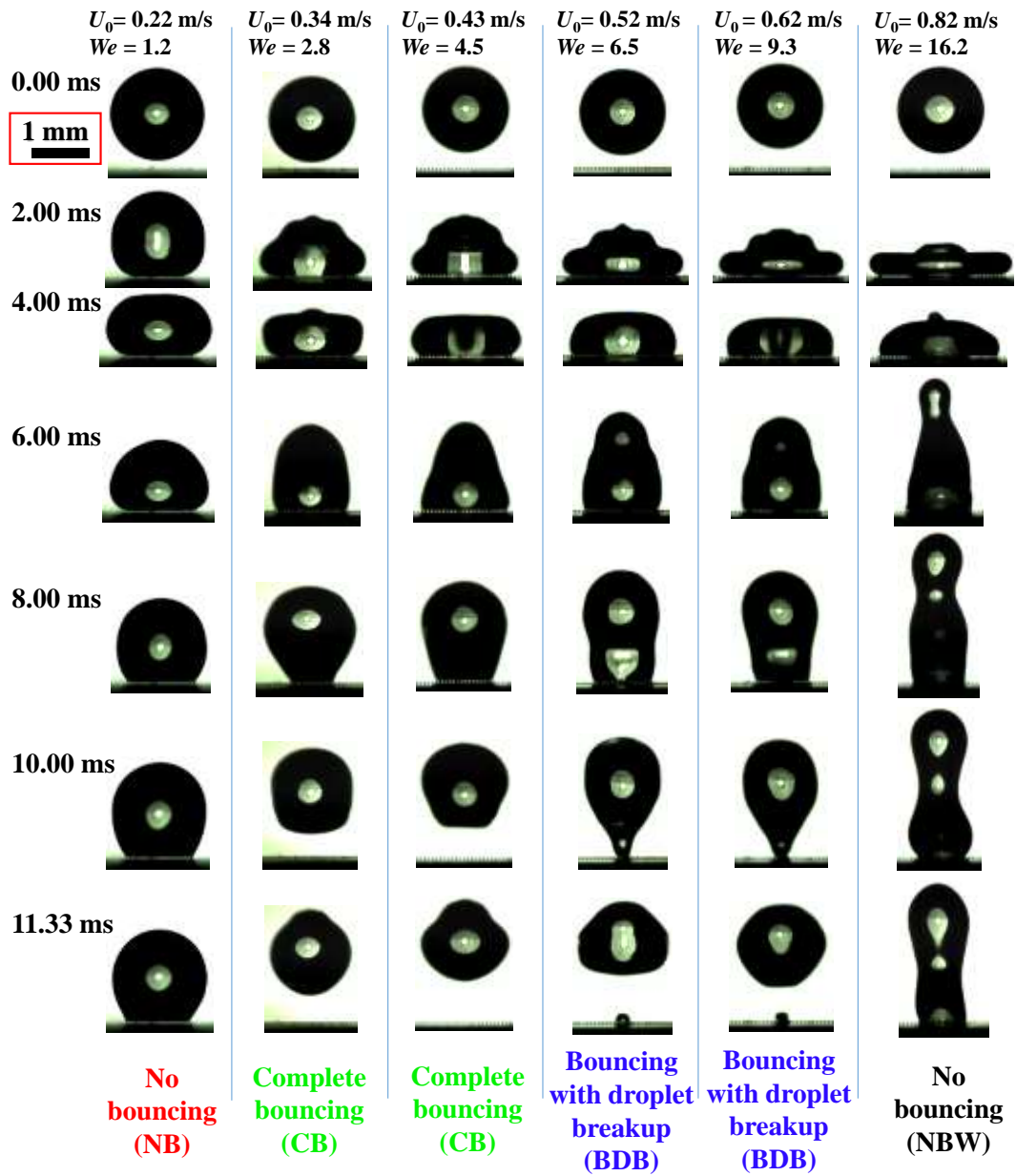

Figure 10: Image sequence obtained by high-speed visualization in transverse direction during impact of a microliter water droplet of 1.7 mm diameter on a microgrooved surface with $p = 76$ µm. Columns show different cases of Weber number (or impact velocity) and scale is shown on top-left. See also the associated supplementary movie, S2.avi.

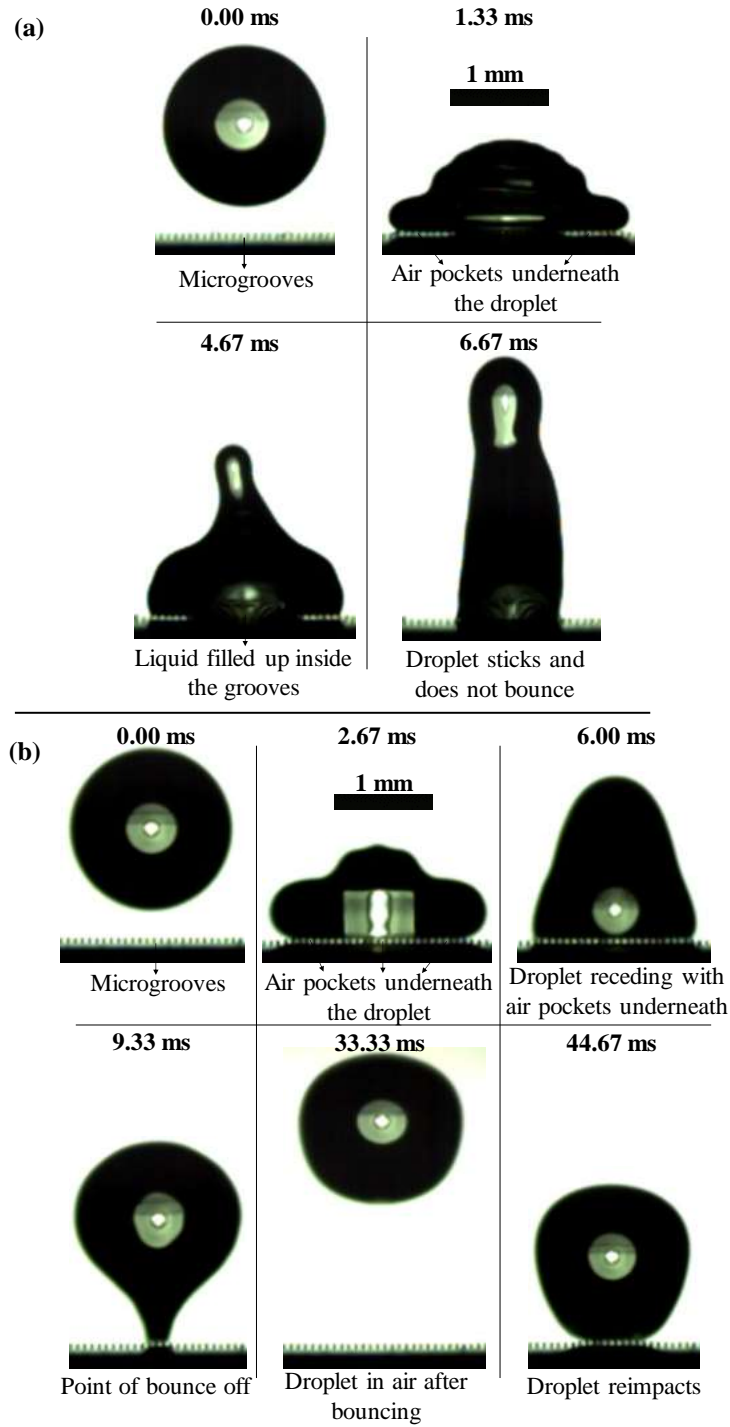

Figure 11: Image sequence obtained by high-speed visualization in transverse direction during impact of 1.7 mm water droplet on a microgrooved surface with pitch of 76 µm. (a) Droplet impacts at Weber number, $We$ = 16.2 ($U_0$ = 0.82 m/s). Images show the penetration of water into the microgrooves i.e. Cassie to Wenzel wetting transition (NBW) (b) Droplet impacts at Weber number, $We$ = 4.5 ($U_0$ = 0.43 m/s) and bounces off the surface (CB). Water does not penetrate into the grooves during spreading or receding and the droplet rests on the ridges after it impacts again the surface.

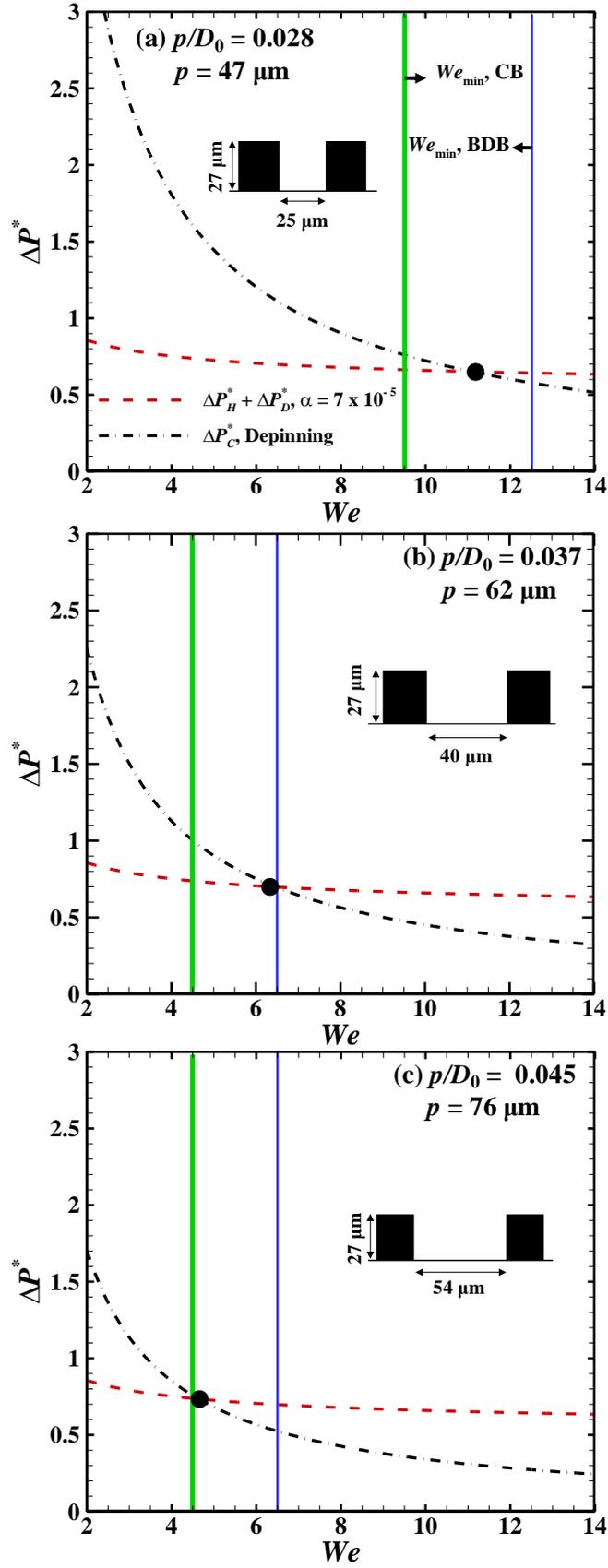

Figure 12: Estimation of theoretical critical Weber number (*We*) by balancing dimensionless capillary pressure ($\Delta P_C^*$) calculated for depinning mechanism with sum of dimensionless

hammer ($\Delta P_H^*$) and dynamic pressure ($\Delta P_D^*$). The critical Weber number is represented by a filled circle and obtained by intersection of plots of $\Delta P_C^*$ and $\Delta P_H^* + \Delta P_D^*$. The minimum Weber numbers obtained in measurements for CB and BDB are shown by vertical lines. Three cases of the pitch are considered: (a) $p = 47$ μm ($p/D_0 = 0.028$), (b) $p = 62$ μm ($p/D_0 = 0.037$) and (c) $p = 76$ μm ($p/D_0 = 0.045$).

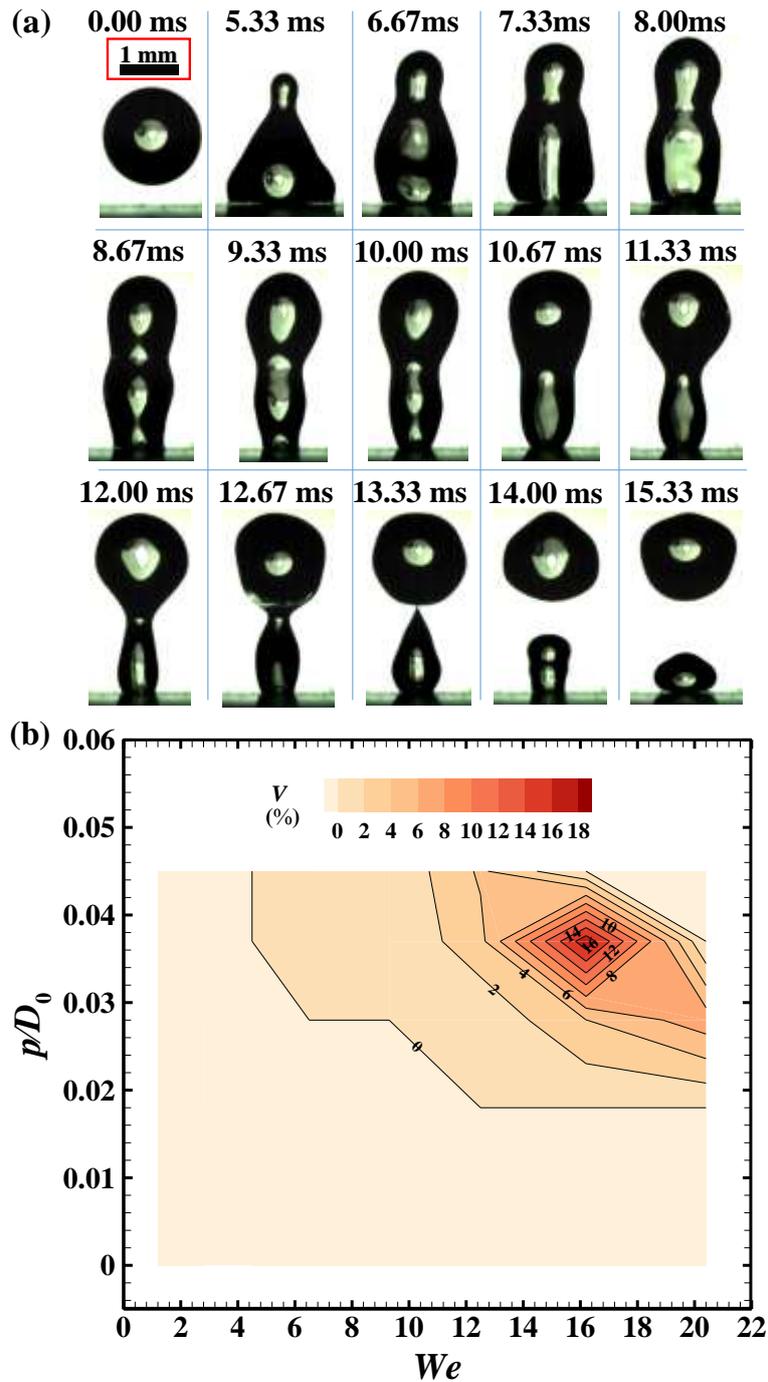

Figure 13: (a) Image sequence obtained by high-speed visualization in transverse direction of impact of 1.7 mm diameter water droplet on a microgrooved surface with of pitch, $p = 62$ µm and at Weber number, $We = 16.2$ ($U_0 = 0.82$ m/s). Scale is shown on top-left. (b) Contour of volume of daughter droplet left on the surface after droplet bouncing and breakup (BDB) on $We$-$p/D_0$ plane.

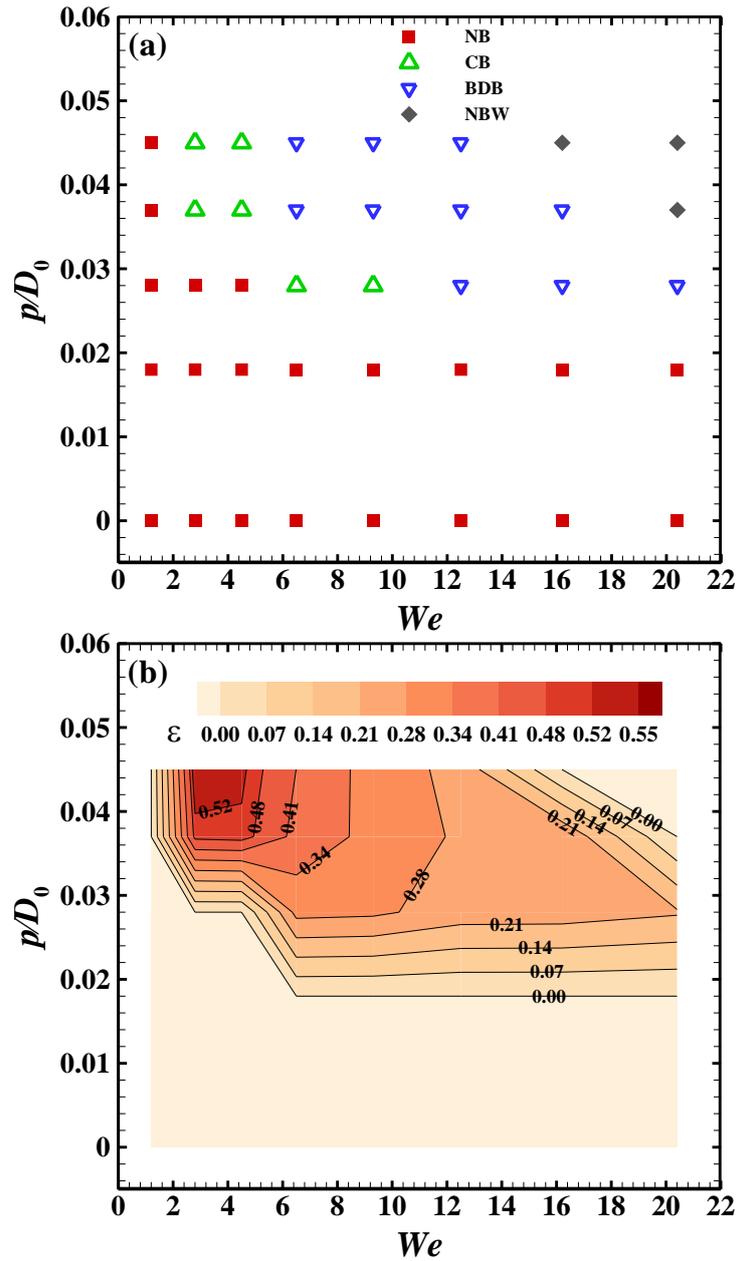

Figure 14: (a) Regime map on Weber number (*We*) – dimensionless pitch (*p/D*₀) plane for different impact outcomes obtained in the present study for 1.7 mm water droplet. The outcomes are no bouncing (NB), complete bouncing (CB), bouncing with droplet breakup (BDB) and no bouncing due to Cassie to Wenzel wetting transition (NBW). (b) Contour of coefficient of restitution for CB and BDB on *We* - *p/D₀* plane.

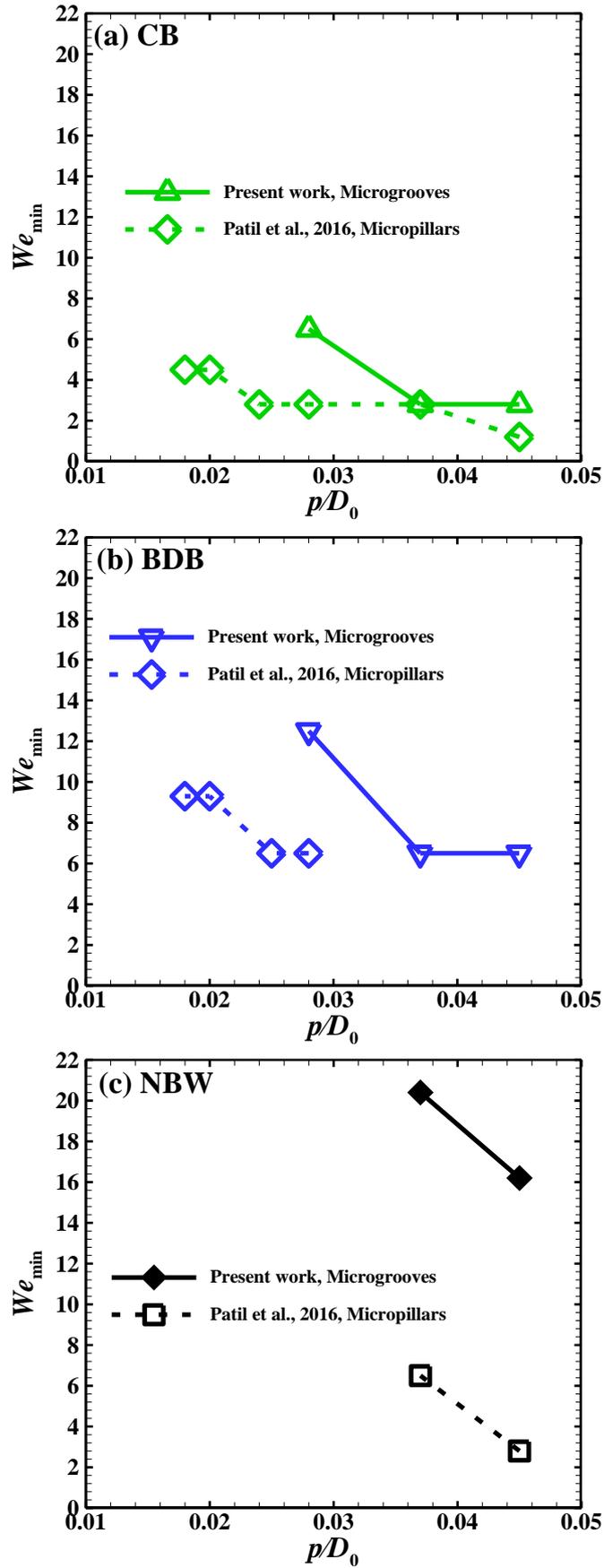

Figure 15: Comparison between $We_{min}$ required for different outcomes, CB (a), BDB (b), and NBW (c) on microgrooved and micropillared surfaces for different cases of pitch.

# 11 TOC graphic

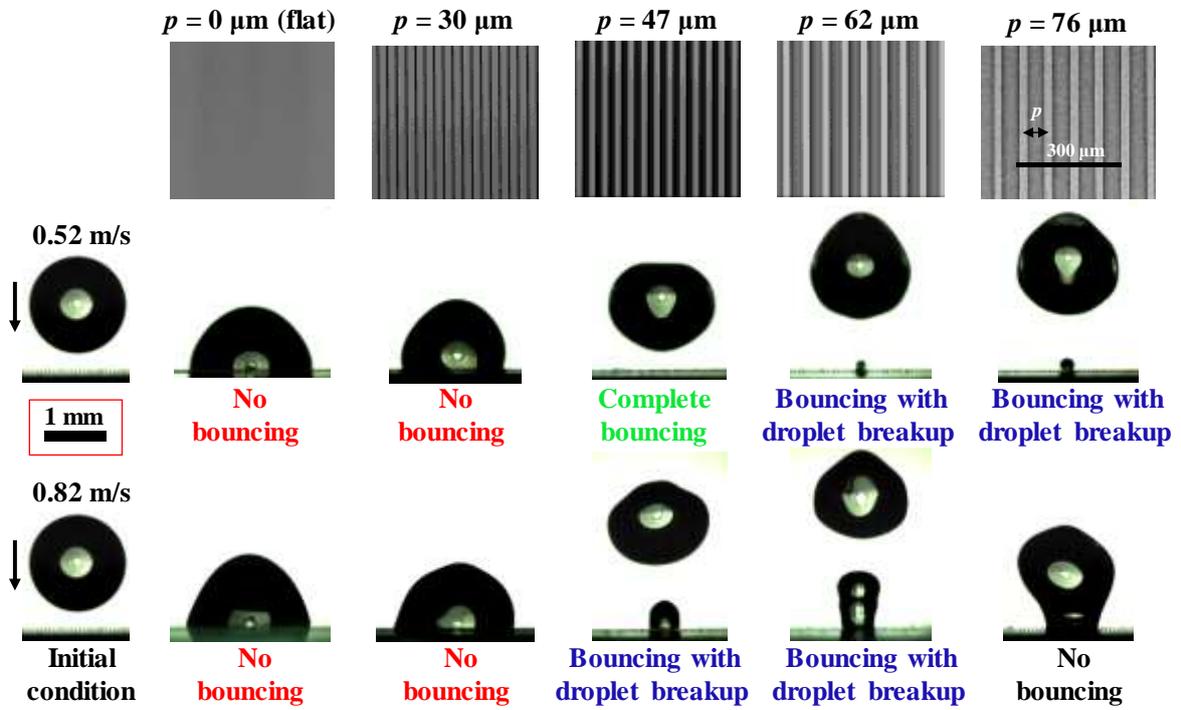